\documentclass[11pt]{article}
\pdfoutput=1
\usepackage{graphicx}
\usepackage{subfigure,amssymb,amsmath}
\usepackage{cite,color,xcolor,url,cancel,ulem}
\usepackage{bm}
\usepackage{float}
\usepackage{jheppub}
\definecolor{light-gray}{gray}{0.75}

\def\displ{{}}

\def\Vevs{{\it vevs}}
\def\321{$\rm SU(3)_C\times SU(2)_L\times U(1)_Y$}

\def\10{SO(10)}

\def\sb{\tilde{b}}

\def\lspone{\widetilde\chi_1^0}
\def\mlspone{m_{\lspone}}

\def\sigmachionesip{\sigma_{\lspone p}^{SI}}

\def\gsim{\ ^>\llap{$_\sim$}\ } 
\def\amususy{a_\mu^{\rm SUSY}}
\def\gmin2{(g-2)_\mu}
\def\bsg{{\rm Br}(B \rightarrow X_s +\gamma)}
\def\bsmumu{{\rm Br}(B_s \rightarrow \mu^+ \mu^-)} 

\def\issue(#1,#2,#3){{\bf #1}, #2 (#3)}
\def\PREP(#1,#2,#3){Phys.\ Rep. \issue(#1,#2,#3)}

\newcommand{\ms}[2]{m_{\tilde{#1}_{#2}}}

\newcommand{\MW}{M_{W}}

\newcommand{\gev}{\rm GeV}
\newcommand{\tev}{\rm TeV}
\def\lspone{\widetilde\chi_1^0}
\def\mlspone{m_{\lspone}}

\title{
Probing Non-holomorphic MSSM via precision constraints, dark matter and LHC 
data}

\author[a]{Utpal Chattopadhyay,}
\author[b]{Abhishek Dey}

\affiliation[a]{Department of Theoretical Physics, Indian Association 
for the Cultivation of Science,\\ 
2A \& B Raja S.C. Mullick Road, Jadavpur, 
Kolkata 700 032, India}
\affiliation[b]{Maulana Azad College, Government of West Bengal,\\ 8 Rafi Ahmed Kidwai Road, Kolkata 700013, India
	}
\emailAdd{tpuc@iacs.res.in}
\emailAdd{dey.abhishek111@gmail.com}

\abstract{
In this analysis we explore the phenomenological constraints of models 
with non-holomorphic soft SUSY 
breaking terms in a beyond the MSSM scenario 
having identical particle content. The model referred  as NHSSM shows various 
promising features like the possibility of a strong reduction in electroweak fine-tuning even for a scenario of a heavy 
higgsino type of LSP, a fact that is unavailable in pMSSM models. The other 
important aspect is satisfying the muon $g-2$ data even for a small 
$\tan\beta$ via a small value of coupling $A_\mu'$ 
associated with the tri-linear non-holomorphic soft term. Thus, a 
large SUSY contribution to muon $g-2$ is possible  
even for a significantly large smuon mass $m_{\tilde {\mu_1}}$. 
The Higgs mass radiative corrections are contributed by both the 
holomorphic and non-holomorphic trilinear soft parameters 
$A_t$ and $A_t'$, thus diluting the requirement to have a larger $A_t$ 
to satisfy the Higgs mass data.  The model 
also provides with valid parameter space 
satisfying the constraint of $\bsg$ for large values of 
$\tan\beta$, a scenario unfavourable in pMSSM.
}  
\begin{document}
\maketitle
\section{Introduction}
The discovery of the Higgs Boson at the ATLAS\cite{HiggsDiscoveryJuly2012} and the CMS
\cite{Chatrchyan:2013lba} experiments of the Large Hadron Collider (LHC) marks the completion
of particle searches within the realm 
of the Standard Model (SM)\cite{SMrefs}.  The SM is quite 
successful in explaining electroweak and strong 
interactions and the associated Higgs mechanism is found to be a viable method for generating masses for 
fermions and electroweak gauge bosons.  Despite its 
success in explaining most of the observed experimental 
results, there are many theoretical issues and 
experimental facts that cannot be addressed while staying 
within the SM. The gauge 
hierarchy problem, baryogenesis, the fact that
 neutrinos have masses, the absence of a dark matter candidate, are a 
few of the important issues that 
motivate us to explore Beyond the SM (BSM) scenarios.   
Models involving 
Supersymmetry (SUSY) such as the Minimal Supersymmetric Standard Model 
(MSSM)\cite{SUSYreviews1,SUSYreviews2,SUSYbook1,SUSYbook2} 
are prominent candidates for BSM physics.  However,  
the fact remains that even after the first few years of running of the Large 
Hadron Collider (LHC), SUSY is yet to be found.  This has 
obviously put serious constraints on
various models of low energy SUSY. In the post-Higgs discovery years,  
the lighter Higgs boson of MSSM 
to have a mass of $m_h\sim$~125~GeV\cite{Aad:2015zhl} 
translates into large radiative corrections\cite{djouadi}. 
This demands a heavier top squark sector.  A large fine-tuning is to 
be accepted.  Furthermore, LHC has pushed up the lower limits 
of masses of the first two generations of squarks as well as 
gluino beyond a TeV.  At the same time, SUSY models are increasingly being 
constrained 
via B-physics related measurements at LHCb.  On the other hand, 
regarding dark matter (DM)\cite{dark_matter_refs}, 
the measurements from WMAP/PLANCK\cite{Hinshaw:2012aka,Ade:2013zuv} for the DM relic density or 
LUX\cite{Akerib:2013tjd} experiment for DM direct detection have put significant limits.   
We further emphasize that, the data from the Brookhaven experiment for the anomalous magnetic moment of muon or  
$\gmin2$ points out a significant deviation (3.2$\sigma$) from its SM 
based evaluation while we note that various uncertainties 
of the SM contributions to $\gmin2$ are being reduced over the last few years.    
This leads to stringent  
constraints on the scalar and gaugino sectors of SUSY models.    
A combined 
requirement for 
satisfying the relic density range from WMAP or PLANCK experiments 
apart from satisfying the LHC derived sparticle mass\cite{Agashe:2014kda} bounds particularly 
creates tension so as to have a reasonably large $\amususy$.   Here, $\amususy$
refers to the SUSY contribution to the theoretical evaluation of 
the muon anomaly $a_\mu \equiv \frac{1}{2}{(g-2)}_\mu$ 
which is supposed to be equal to the difference between the experimental value 
and the SM evaluation of the observable.   
The Higgsinos and Wino are typically required to be 
heavy for becoming candidates of 
dark matter.  On the other hand, satisfying $\gmin2$ with dominant contributions from the chargino-sneutrino loops demands non-decoupling higgsinos or wino 
along with a light sneutrino ${\tilde \nu}_\mu $.  We emphasize that 
in models with gaugino mass 
universality like minimal supergravity (mSUGRA)\cite{msugra}/constrained MSSM (CMSSM)\cite{SUSYbook1}, 
the chargino-sneutrino loops 
dominate in $\amususy$\cite{Jegerlehner:2009ry,muong1}.  Thus in the present scenario of sparticle mass limits $\amususy$ is not large enough to explain the observed deviation.  In the parameter space of MSSM that is consistent with the Higgs mass data, 
dark matter relic density, collider limits for scalar and gaugino 
(electroweakino) masses, 
it turns out that the primary contribution to $\amususy$ 
comes from the loop diagrams containing neutralinos and smuons, particularly 
from the L-R mixing terms that scale with $M_1 \mu\tan\beta$\cite{Endo_muong}. 
In order to have a larger $\amususy$ so as to account for 
the deviation the above quantity needs to be large or in other words 
this restricts the smuon mass $m_{{\tilde \mu}_1}$ to become large.  
A comprehensive analysis in a model that identifies the valid region of 
parameter space satisfying limits from flavor physics such as 
that from $\bsg$ and $\bsmumu$ (which we would collectively refer as B-physics 
constraints), dark matter constraints, while also having a moderate degree of  
fine-tuning and most importantly that would easily accommodate the 
$\gmin2$ limits even for a small $\tan\beta$ all at one go, is undoubtedly  
important.  In this analysis, keeping ourselves contained within the 
MSSM particle setup we would like to explore 
whether a consideration of non-holomorphic (NH) soft SUSY breaking 
terms may be able to reduce the stringency arising out of the $\gmin2$ 
constraint in particular apart from satisfying 
all the above mentioned phenomenological requirements. 

 
Away from MSSM, particularly in models with singlet scalars   
NH soft breaking terms potentially fall 
in the class of terms that may cause hard SUSY 
breaking\cite{grisaru,Martin:1999hc,Haber:2007dj,Bagger:1993ji,Ellwanger:1983mg,Jack:1999ud}. Considering, for example, 
a hidden sector SUSY breaking scenario like supergravity, one may generically
consider a spontaneous SUSY 
  breaking due by the vacuum expectation value of an auxiliary field $F$
  belonging to a chiral superfield $X$.
This causes appearance of soft terms in the 
Lagrangian that are associated with the coupling of $X$ 
with another chiral superfield
$\Phi$ or a gauge field strength superfield $W_\alpha^a$.
In a supergravity
  framework where the mass scale $M$ is large (typically the Planck mass)
  one obtains the following \cite{Martin:1999hc}. 
  \begin{equation}
    -{\cal L} =
    \left ({1\over M} [X W^{\alpha a} W^a_\alpha]_F
    + {1\over M}[X \Phi^3]_F
    + {\mu\over M}[X \Phi^2]_F
    \right ) + {\rm c.c.} +
           {1\over M^2} [X^* X \Phi^* \Phi]_D .
           \label{softorigin}   
  \end{equation}
  Here, the parameter $\mu$ is introduced for the soft term so as to
  follow closely with the usual MSSM notation\cite{SUSYbook1}.
  Considering the vacuum expectation value 
  $\langle X \rangle = \theta\theta <F>$ and denoting $<F>$ simply by $F$ 
 one has the usual soft terms of MSSM namely the gaugino mass
  term, the cubic and the analytic scalar squared mass terms and a
  non-analytic scalar mass term coming out of the above D-term contribution
  as given below. 
  \begin{equation}
  -{\cal L} = \left (
  {F\over M} \lambda^a \lambda^a
  + {F\over M} \phi^3
  + {\mu F\over M} \phi^2\right )
  + {\rm c.c.} +
  {|F|^2 \over M^2} \phi^* \phi .
  \end{equation}
We note that $F/M$ should refer to a weak scale mass 
which we consider here as the W-boson mass $M_W$. Apart from the above 
contributions,  
       there can be other D-term contributions that
      are classified as ``maybe soft''\cite{Martin:1999hc} such as  
      the ones arising out of 
$ {1\over M^3}[ X X^* \Phi^2 \Phi^* ]_D$ and  
      $ {1\over M^3}[ X X^* D^\alpha \Phi D_\alpha \Phi ]_D$ that give
      rise to NH
      terms in the Lagrangian like $\phi^2 \phi^* $ and $\psi\psi $ both with
      coefficients $\displ {|  F  |^2 \over M^3} \,\sim\,
      {M_W^2 \over M} $. These terms in a broader sense 
      may 
      cause quadratic divergence, thus become hard SUSY breaking terms,  
in scenarios where the visible 
      sector contains a singlet superfield. Nevertheless,   
      the terms are generally 
      highly suppressed in a supergravity type of scenario.
      Indeed this is why such 
      NH contributions to the
      Lagrangian are traditionally ignored while discussing models with
      high scale based SUSY breaking. 
      We must, however, note that, as pointed out in 
\cite{Frere:1999uv}, in absence of any  
      gauge singlet field in the visible sector such 
      suppression may not be possible if the supersymmetry breaking effect
      is communicated to the visible sector at a lower energy. 
For example, this may happen in scenarios with gauge-mediated 
supersymmetry breaking\cite{Frere:1999uv}\footnote{For scenarios with 
$F$-term SUSY breaking leading to $\frac{1}{M^3}$ type of suppression in 
nonstandard supersymmetry breaking terms including Dirac gaugino mass 
terms see ref.\cite{Martin:2015eca}.   
}.  Thus having no gauge singlet superfields MSSM may as well 
      include such NH terms that are
      soft SUSY breaking in nature.  The terms can hardly be ignored in 
the most general sense.   This was  
discussed or at least pointed out in several 
works\cite{Martin:1999hc,Haber:2007dj,Bagger:1993ji,Jack:1999ud,Jack:1999fa,Hetherington:2001bk,Jack:2004dv,Cakir:2005hd,Un:2014afa1}.  

  Apart from the references shown above related to the issue of 
absence of quadratic divergence in MSSM in presence of NH soft terms and 
the possible origin of the terms in relation to a hidden sector 
SUSY breaking model, we will now briefly refer to a 
few specific works related to phenomenology.  A general analysis with 
NH soft terms 
that incorporated renormalization group evolutions  
with or without using R-parity violation was 
presented in Ref.\cite{Jack:2004dv}. This was followed by 
a study\cite{Cakir:2005hd} with NH SUSY breaking terms in an essentially
Constrained Minimal Supersymmetric Model (CMSSM) setup with all input
parameters including the NH ones being given at the gauge coupling unification scale while exploring relevant phenomenological constraints like $\bsg$. 
Analyses that particularly  
focussed on the Higgs sector and constraints like $\bsg$ 
are Refs.\cite{Un:2014afa1,Un:2014afa2}.  However, in contrast to  
Ref.\cite{Un:2014afa1} that used input parameters given entirely at 
the unification scale, Ref.\cite{Un:2014afa2} analyzed in a mixed set-up 
where the NH parameters were given at the electroweak  
scale in an otherwise CMSSM type of setup. 
Similar mixed input parameters were used  
in Ref.\cite{Hetherington:2001bk} that discussed
reparametrization invariance in special circumstances of the choice 
of parameters\footnote{This, however, does not apply to our analysis 
involving unconstrained NH SUSY breaking parameters.}, spectra as well as the effect on     
fine-tuning. 
Ref.\cite{Solmaz2009} may be seen for its emphasis on the Higgs 
sector while considering CP-violating phases.  
Nonstandard SUSY breaking was also used 
in phenomenological studies with R-parity violating 
NH soft SUSY breaking terms 
in MSSM framework in Refs.\cite{Jack:2004dv,Hambye:2000zs,SRoy}. 



Unlike all the previous analyses where universal models of CMSSM type 
were considered 
with NH parameters being given either at the unification scale or at the 
electroweak scale, our work on the  
Non-Holomorphic Supersymmetric Standard Model (NHSSM)   
will entirely use electroweak scale input parameters 
similar to what is considered in phenomenological MSSM (pMSSM) 
model\cite{Djouadi:1998di} in relation to MSSM.  
In this phenomenological NHSSM (pNHSSM) framework,  
we will explore the extent NH parameters influence on satisfying 
the $\gmin2$ constraint  apart from the effect on 
electroweak fine-tuning and $m_h$ via the associated radiative 
corrections. 
The Higgs mass limit is achieved for smaller values of $|A_t|$, the 
trilinear coupling parameter corresponding to the top-quark in 
comparison with what is required for MSSM. 
Furthermore, as we will see such NH parameters 
do not affect the fine-tuning measure since the Higgs scalar 
potential would not have any dependence on such parameters\footnote{See Ref.\cite{Ross:2016pml} that appears 
	while this work was being done in which the authors included the 
	NH soft terms in their discussion on Higgsino dark matter while extending MSSM. Our 
	work in NHSSM 
	involves electroweak scale input parameters for all the 
soft terms similar to pMSSM 
	unlike their analysis with RG evolutions and   
	unification scale input parameters.}.

We will additionally focus on 
the low energy processes like $\bsg$ and $\bsmumu$
that also receive a significant amount of  contributions from the NH terms. 
Particularly these constraints may indeed be quite severe for large values of $\tan\beta$ in MSSM.
  However, the NH terms are able to  
alter the above branching ratios 
and a large region of parameter space that 
would be excluded in MSSM is restored. 
 We will further see that there are 
valid regions of pNHSSM parameter 
space that is consistent with correct relic abundance 
for dark matter (DM) and these may be probed for 
the Direct and Indirect Detection limits of DM.

Our paper is organized as follows. In Sec.\ref{nhssm_theory} we 
discuss NHSSM and particularly explain the impact of NH 
terms on the i) Higgs and other scalar sectors, ii) charginos and neutralinos, electroweak fine-tuning and  iii) phenomenological aspects related to the constraints coming from 
dark matter, muon $g-2$ as well as $\bsg$ and $\bsmumu$.
We present the results of our analysis in Sec.\ref{results}. 
Finally, we conclude in Sec.\ref{conclusion}.

\section{Non-Holomorphic Supersymmetric Standard Model}
\label{nhssm_theory}
We remind that MSSM is considered to have only holomorphic soft 
SUSY breaking terms. The trilinear soft terms, in particular,  
are given by as follows \cite{SUSYbook1}
\begin{equation}
\label{h_lagrangian}
 -\mathcal{L}_{soft}\supset \tilde{Q}\cdot H_uA_t\tilde{U} + 
 \tilde{U}\cdot H_dA_b\tilde{D} +\tilde{L}\cdot H_d A_{\tau}\tilde{E} +h.c.
\end{equation}
We have only shown here the dominant terms involving the 
third generations of fermions.
It was shown that in
the absence of any Standard Model gauge singlet it is possible to extend the SUSY 
breaking soft sector by including NH soft SUSY 
breaking terms, without 
aggravating any quadratic divergence \cite{grisaru,Bagger:1993ji}.  
Thus the NH soft terms 
of the NHSSM in general that include trilinear coupling terms 
as well as a coupling term involving Higgsinos are 
given by\cite{Un:2014afa1,Un:2014afa2},
\begin{equation}
\label{nh_lagrangian}
 -\mathcal{L'}_{soft}\supset \tilde{Q}\cdot H_{d}^c A_{t}'\tilde{U} + 
 \tilde{U}\cdot H_{u}^c A_{b}'\tilde{D} +\tilde{L}\cdot H_{u}^c A_{\tau}'\tilde{E} +  
 \mu '\tilde{H_u}\cdot \tilde{H_d} +h.c.
\end{equation}

We will now discuss the effect of involving NH terms 
on several sectors of MSSM sparticle spectra, particularly in the 
Higgs sector, the squark and slepton sectors, the electroweakinos 
(charginos and neutralinos) apart from its effect on a fine-tuning
measure. We will also include low energy data from 
precision experiments like $\gmin2$\cite{Bennett:2006fi,Roberts:2010cj,Nyffeler:2013lia,Hagiwara:2011af}, $\bsg$\cite{Amhis:2012bh}, $\bsmumu$\cite{Chatrchyan:2013bka,Aaij:2013aka,CMSandLHCbCollaborations:2013pla}
and cosmological 
observables like dark matter relic density 
while also taking into account the LHC bounds for 
sparticles and the Higgs mass data\cite{Aad:2015zhl}. 

\subsection{Influence of non-holomorphic terms on the scalar sector}
\label{nhssm_scalar}
The NH trilinear coupling parameters may cause a significant amount of change 
in the masses of squarks and sleptons. For example, 
the mass matrix for the up type of scalar quark, in general, is 
given by \cite{Un:2014afa1,Un:2014afa2}. 
\begin{eqnarray}
\label{stop_mass}
 M_{\tilde{u}}^2=&\left(\begin{matrix}
   m_{\tilde{Q}}^2+(\frac{1}{2}-\frac{2}{3}\sin^2\theta_{W})M_Z^2\cos2\beta +m_u^2&\hspace{2mm}
   -m_u(A_u-(\mu+A_u')\cot\beta) \\
   -m_u(A_u-(\mu+A_u')\cot\beta)  & \hspace{-2mm}
   m_{\tilde{u}}^2+\frac{2}{3}\sin^2\theta_{W} M_Z^2 \cos2\beta +m_u^2 
                    \end{matrix} \right).\hspace{6mm}
\label{stopmassmatrix}
\end{eqnarray}
Similar matrices for a slepton or a down-type of squark may be written  
as given below. 
\begin{eqnarray}
\label{slepton_mass}
M_{\tilde{e}}^2=&\left(\begin{matrix}
M_{\tilde{l_L}^2}+M_z^2(T_{3L}^{\tilde{e}}-Q_e\sin^2\theta_{W})\cos2\beta +m_e^2 &\hspace{1mm}
-m_e(A_e-(\mu+A_e')\tan\beta) \\
\hspace{-3mm} -(A_e-(\mu+A_e')\tan\beta) m_e & \hspace{-6mm}
M_{\tilde{l_R}^2}+M_z^2 Q_e\sin^2\theta_{W}\cos2\beta +m_e^2
\end{matrix} \right).\hspace{6mm}
\end{eqnarray}
Clearly, $\mu$ of MSSM that contributes to L-R mixing of squark is 
replaced by $\mu+A_u'$ in NHSSM.  
The contributions of the NH 
terms will thus be more effective for i) low $\tan\beta$ in the 
case of up type 
of  squarks and ii) large $\tan\beta$ in case of down type of squarks or 
sleptons.
\\

\noindent
An effect on top-squark sector is transmitted to the  
Higgs mass radiative corrections.  We remind that in the MSSM framework the 
discovery of the Higgs boson with a mass of 
$125.09 \pm 0.24 \gev$\cite{Aad:2015zhl} is translated into  
a large radiative corrections 
to the mass of the lighter neutral CP-even Higgs 
boson $h$.\footnote
{Precision measurements indicate that the discovered Higgs 
Boson is consistent with being SM-like\cite{Aad:2015zhl}. 
With the above in perspective we consider the decoupling regime of MSSM 
Higgs characterized by $M_Z^2 \ll M_A^2$\cite{djouadi}. 
}. The above requirement pushes up 
the masses of the top-squarks in MSSM which 
in turn indicates the need for a large value of $|A_t|$ so as to have a
larger Left-Right mixing.

\noindent
Considering the top-stop loops which constitute the most 
contributing terms, the 
above radiative corrections in NHSSM read as follows. 
\begin{equation}
\label{higgs_loop}
\Delta m_{h,top}^2= \frac{3 g_2^2 {\bar m}_t^4}{8 \pi^2 \MW^2} 
\left[\ln\left(\frac{\ms{t}{1} \ms{t}{2}}{{\bar m}_t^2}\right) + \frac{X_t^2}{\ms{t}{1}\ms{t}{2}} 
\left(1 - \frac{X_t^2}{12\ms{t}{1}\ms{t}{2}} \right) \right]. 
\end{equation}
Here, $X_t=A_t-(\mu+A_t')\cot\beta$.  Clearly $A_t'=0$ corresponds to 
the MSSM result.  Here ${\bar m}_t$ refers to the running top-quark mass 
that includes corrections from the electroweak, QCD and SUSY QCD effects.  
The maximal mixing scenario refers to 
$X_t={\sqrt 6} M_S$\cite{SUSYbook1,SUSYbook2} where $M_S=\sqrt{\ms{t}{1}\ms{t}{2}}$. 
We note that unlike MSSM, with a suitable combination of the signs 
of $A_t$ and $A_t'$, it is possible to limit $|A_t|$ so as to satisfy the Higgs mass constraint.

We shall now discuss the effect of considering NHSSM  
on the Electroweakino (Chargino-Neutralino) sector and  
fine tuning.

\subsection{Electroweakinos in NHSSM}
\label{Electroweakinos}
The NH parameter $\mu'$ modifies both the chargino 
and neutralino mass matrices. Essentially the charginos and neutralinos 
now have higgsino components corresponding to a higgsino mass of 
$|\mu-\mu'|$. Thus in NHSSM the neutralino mass matrix reads\cite{Un:2014afa1,Un:2014afa2}:
\begin{eqnarray}
\label{neutralino_mass}
 M_{\widetilde{\chi^0}}=&\left(\begin{matrix}
                        M_1 & 0 & -M_Z\cos\beta \sin\theta_W & M_Z\sin\beta \sin\theta_W \\
                        0 & M_2 & M_Z\cos\beta \cos\theta_W & -M_Z\sin\beta \cos\theta_W \\
                       -M_Z\cos\beta \sin\theta_W & M_Z\cos\beta \cos\theta_W & 0 & -(\mu-\mu')\\
                        M_Z\sin\beta \sin\theta_W & -M_Z\sin\beta \cos\theta_W & -(\mu-\mu') & 0
                                           \end{matrix} \right).\hspace{6mm}
\end{eqnarray}
Similarly for the chargino matrix we have\cite{Un:2014afa1,Un:2014afa2}:
\begin{eqnarray}
\label{chargino_mass}
  M_{\widetilde{\chi^{\pm}}}=&\left(\begin{matrix}
                                 M_2 & \sqrt{2} M_W\sin\beta \\
                                 \sqrt{2} M_W\cos\beta & -(\mu-\mu')\\
                                \end{matrix}\right).
\end{eqnarray}
We note that the LEP bound on the lighter chargino mass will essentially 
apply to $|\mu-\mu'|$ instead of $\mu$.  However, as we will see below the fine-tuning measure is still dependent on $\mu$ rather than $\mu'$ and 
this provides with a quite unique signature of a possibility of having 
low fine-tuning irrespective of the nature of dark matter considered 
in NHSSM. For example, one can have a bino-like LSP even for very 
low $\mu$. 

\subsection{Electroweak fine tuning in pNHSSM}
\label{ewft_theory}
The nonholomorphic trilinear parameters are 
associated with charged or colored scalars whereas the parameter $\mu'$
is associated with fermions (higgsinos). Thus with no influence of the 
nonholomorphic soft breaking terms 
on the neutral scalar potential, the latter is same as that of MSSM as given 
below.  
\begin{equation}
\label{higgs_potential}
 V=(m_{H_u}^2 + \mu^2)|H_u^0|^2 + (m_{H_d}^2 + \mu^2)|H_d^0|^2 -b(H_u^0H_d^0 +h.c.)+\frac{1}{8}(g^2+g'^2)(|H_u^0|^2-|H_d^0|^2)^2, 
\end{equation}
where $\mu$ is the bilinear Higgs 
mixing parameter of the superpotential, $m_{H_u}$, $m_{H_d}$ 
are scalar mass parameters and $b$ is the Higgs mixing parameter 
within the SUSY breaking soft sector.
After minimization with respect to the vacuum expectation values
(\Vevs)~ of the neutral Higgs scalars, one finds,  
the well-known relations for electroweak symmetry breaking
\cite{SUSYbook1,SUSYbook2}
\begin{equation}
\label{rewsb1}
\frac{m_Z^2}{2}=\frac{m_{H_d}^2-m_{H_u}^2\tan^2\beta}{\tan^2\beta-1}-{|\mu|}^2,
\end{equation}
and,
\begin{equation}
\label{rewsb2}
\sin{2\beta}=\frac{2b}{m_{H_d}^2+m_{H_u}^2+2{|\mu|}^2}.
\end{equation}
 Clearly, at the tree-level one requires a fine cancellation 
between the two terms in the right-hand side of Eq.\ref{rewsb1} 
coming from SUSY breaking 
parameters $m_{H_d},m_{H_u}$
and supersymmetry preserving parameter $\mu$ 
for obtaining the left-hand side namely $m_Z^2/2$ 
where $m_Z$ refers to the measured value of 91.2~\gev.~ 
The degree of cancellation broadly indicates a measure of the fine-tuning at the 
electroweak scale.  We consider here a general definition  of electroweak fine tuning (EWFT) that uses log derivatives\cite{ewft_papers,baer_ewft,ewftotherworks}, namely 
\begin{equation}
 \Delta_{p_i}=\left|\frac{\partial \ln m_Z^2(p_i)}{\partial\ln p_i}\right|,
\end{equation}
where $p_i \equiv\{\mu^2,b,m_{H_u},m_{H_d}\} $ are the parameters 
that determine the tree-level value of $Z$ boson mass.  
The total EWFT at low scale is given by, 
\begin{equation}
 \Delta_{Total}=\sqrt{\sum\nolimits_{i}\Delta_{p_i}^2}.
\label{formulaewft}
\end{equation}
It turns out that with large $\tan\beta$ the most important terms are 
$\Delta(\mu) \simeq\frac{4\mu^2}{m_Z^2}$ and $\Delta(b) \simeq 
\frac{4M_A^2}{m_Z^2\tan^2\beta}$. 
The above expressions show that even for a moderately large $\tan\beta$,  
a small value of EWFT demands a lower value of $\mu$. 
However, very small values of $\mu$ are excluded in MSSM due to LEP bound of 
lighter chargino mass and we will see that this has an important significance 
in relation to the fine tuning in NHSSM in Sec.\ref{results}.
We however point out that for small $\tan\beta$ and
very small $\mu$ (typically much smaller than the above chargino mass limit)
situation may arise where $\Delta {(m_{H_u})}$ and
$\Delta {(m_{H_d})}$ become larger than $\Delta(\mu)$\cite{ewft_3}.
As a result even for
negligible values of $\mu$ one may obtain finite EWFT as we will see in
Sec.\ref{finetuninganalysis}. 

   In our discussion on EWFT we must, however, remember that, we should include the principal corrections due to one-loop 
radiative effects due to top-stop loops and this leads to\cite{ewft_3},  
\begin{equation}
\label{deltaloop}
\delta m_{H_u}^2 = -\frac{3y_t^2}{8\pi^2}(m_{\tilde{t_L}}^2 + m_{\tilde{t_R}}^2 +|A_t^2|)\log\left(\frac{\Lambda}{m_{\tilde{t}}}\right).
\end{equation}
\noindent
Albeit this depends on the choice of the cut-off scale $\Lambda$. 
The requirement of 
a large $A_t$ in the post Higgs@ 125 GeV scenario 
increases $\delta m_{H_u}^2$ which specially shows the need to include the 
one-loop corrections shown above.  This obviously enhances the 
EWFT, although we will not include this 
effect in our EWFT measure. We must, however, point out that minimizing the 
Higgs potential near a scale where the logarithmic term vanish in Eq.\ref{deltaloop} may reduce the 
requirement of inclusion of the one-loop radiative 
corrections significantly as it was discussed in the context of 
hyperbolic branch and electroweak fine-tuning Ref.\cite{ewft_earlier_works}. 

As noted before, a small value of $\mu$ is consistent with smaller 
EWFT. However, the lighter chargino mass bound from LEP 
limits $\mu$ hence to the EWFT measure not to become too small.  In NHSSM the higgsino content of electroweakinos of 
Eqs.\ref{neutralino_mass} and \ref{chargino_mass} depends 
on $|\mu-\mu'|$, thus EWFT may become small irrespective of the 
mass of electroweakinos.   
\subsection{Low Energy Constraints viz. $(B \rightarrow X_s + \gamma)$ and
$(B_s \rightarrow \mu^+ \mu^-)$ in NHSSM}
\label{bphysics}

It is known that rare B-decays within the SM like $(B \rightarrow X_s +\gamma)$, $(B_s \rightarrow \mu^+ \mu^-)$ that are helicity suppressed 
may have large contributions from the radiative corrections 
due to superpartners in the loops.  
The SM contributions to $(B \rightarrow  X_s + \gamma)$ almost saturate
the experimental data. Thus any BSM correction should be low enough to accommodate the 
difference of the SM and the experimental results, which is albeit small. 
SUSY parameter space is thus strongly constrained 
via cancellation of dominantly contributing diagrams, while the contributions 
individually may be large.  One must also remember that 
there are next to leading order (NLO) contributions that can also be quite 
significant, specially for large values of $\tan\beta$.
In the SM the dominant radiative corrections come from $t-W$ loops. In MSSM the 
significantly contributing diagrams involve $t-H^\pm$
and ${\tilde t}-{\tilde \chi}^\pm$ 
loops. The contributions from the former loops share the same sign with the $t-W$ loop contributions of the SM.    
In NHSSM, the soft terms from the NH trilinear coupling 
$A_t'$ and the bilinear higgsino coupling $\mu'$ have significant 
influences on the SUSY diagrams associated with the above flavor related 
processes. In contrast to MSSM, loops involving 
left-right mixing of top-squarks in NHSSM are associated with  
the factor $A_t-(\mu+A_t')\cot\beta$ \cite{Un:2014afa1,Un:2014afa2},
whereas for the contributions involving the higgsino loops, $\mu$ is replaced by the difference $(\mu-\mu')$\cite{Un:2014afa1,Un:2014afa2}. Regarding the constraint from 
${\rm Br}(B_s \rightarrow \mu^+ \mu^-)$ that is typically 
stringent for large $\tan\beta$ and small pseudoscalar mass in MSSM models,  
it turns out that 
 the available parameter space that survives after imposing the 
${\rm Br}(B \rightarrow  X_s + \gamma)$ constraint, 
is not affected much when one imposes the constraint from ${\rm Br}(B_s \rightarrow \mu^+ \mu^-)$. 
\subsection{Muon $g-2$}
\label{intromuong}
The anomalous magnetic moment of muon ($a_\mu=\frac{1}{2}{(g-2)}_\mu$)
is an extremely important constraint for new physics\cite{Jegerlehner:2009ry}. 
The experimental data
($\equiv a_{\mu}^{exp}$)\cite{Bennett:2006fi,Roberts:2010cj}
shows more than 3$\sigma$ level of deviation from the SM 
prediction ($\equiv a_{\mu}^{SM}$)\cite{Hagiwara:2011af,Nyffeler:2013lia}. 
The difference of the two values for a BSM contribution amounts to: 
\begin{align}
\Delta a_{\mu} = a_{\mu}^{exp} - a_{\mu}^{SM} = (29.3\pm 9.0)\times 10^{-10} . 
\end{align}
The above result leads to the following 2$\sigma$ and 1$\sigma$ limits for 
$a_{\mu}^{SUSY}\equiv \Delta a_{\mu}$, where $a_\mu^{SUSY}$ refers to the 
contributions to the muon magnetic moment 
coming from the loop level diagrams involving 
SUSY particles. The limits of $\amususy$ becomes: 
\begin{align}
11.3\times 10^{-10} < a_{\mu}^{SUSY} < 47.3\times 10^{-10} 
\quad (2\sigma) 
\label{sigma2} 
\end{align}
and, 
\begin{align}
20.3\times 10^{-10} < a_{\mu}^{SUSY} < 38.3\times 10^{-10}
\quad (1\sigma).
\label{sigma1} 
\end{align} 
The Feynman diagrams containing chargino-sneutrino
and neutralino-smuon loops produce the most 
dominant SUSY contribution to $a_\mu^{SUSY}$ 
\cite{Jegerlehner:2009ry,muong1}.  
In a bino like LSP scenario with $M_1 < \mu \ll M_2$,
significantly large $a_\mu^{SUSY}$ is achievable
in MSSM via the presence of very light smuon \cite{Endo_muong}.
This is particularly 
true when lighter chargino is not so light or in other words $\mu$ is not very 
small.  However, LHC is increasingly pushing up  
the masses of sleptons and this would require a large value of $\mu$ 
in order to accommodate the muon $g-2$ data\cite{Endo_muong} in MSSM.  A large $\mu$ is obviously not desirable in the context of EWFT.
Moreover, such lower values of $m_{\tilde{\mu_1}}$
may be disfavored by the LHC data.   
Unlike MSSM, $\amususy$ can be enhanced significantly in NHSSM because of additional terms 
proportional to $A_{\mu}'\tan\beta$.  This is true even 
in a very natural scenario ({\it i.e.} with small EWFT) characterized by low $\mu$ 
along with relatively heavier smuons.  

\noindent
\section{Results}
\label{results}
The focus of our analysis in this section would be the important 
features of NHSSM in relation with MSSM, particularly the effects of 
considering NHSSM on    
i) Higgs boson mass, ii) flavor violating 
processes like $\bsg$ and $\bsmumu$ 
iii) SUSY contributions to muon $g-2$, iv) dark matter relic density and v) electroweak 
fine-tuning.  
Regarding the last two points  we will 
particularly demonstrate the fact that NHSSM allows to 
have a higgsino-like DM with a sufficiently low electroweak 
fine-tuning, a feature almost impossible to get in MSSM. We will additionally 
show a few benchmark points consistent with the various 
above constraints.
\subsection{Impact of non-holomorphic soft parameters on $m_h$ }
\label{mh_analysis}
\begin{figure}[htb]
     \begin{center}
        \subfigure[]{%
            \label{fig:Higgs_10}
            \includegraphics[width=0.51\textwidth]{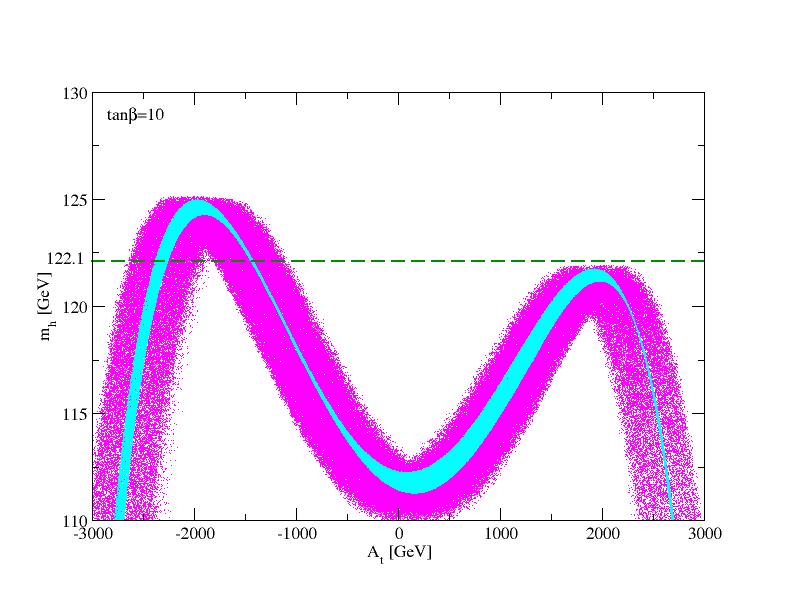}
        }%
        \subfigure[]{%
           \label{fig:Higgs_40}
           \includegraphics[width=0.51\textwidth]{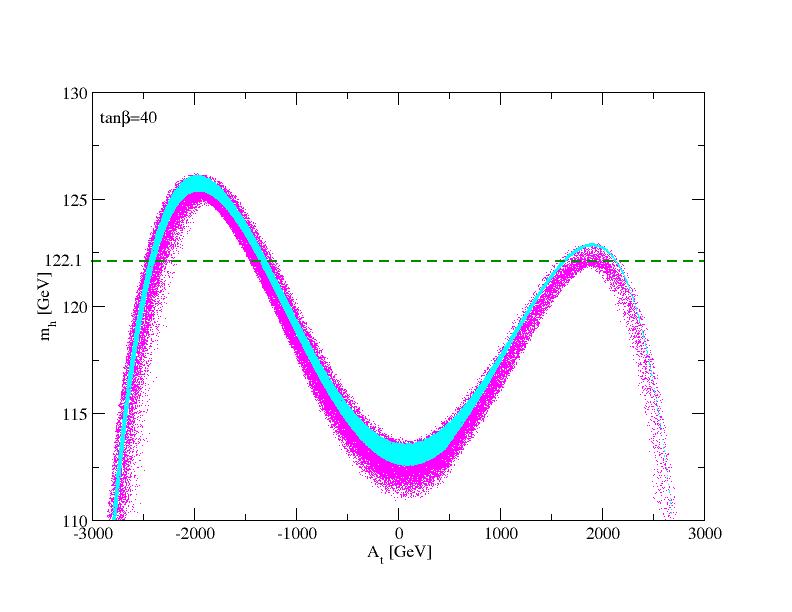}
        }
        \caption{\it Fig.\ref{fig:Higgs_10} shows the variation of $m_h$ against $A_t$ for the 
scanning ranges of Eq.\ref{scan-ranges_10_higgs} for  
$\tan\beta=10$.  The magenta and cyan colored regions correspond to 
NHSSM and MSSM respectively. Fig.\ref{fig:Higgs_40} is same as Fig.\ref{fig:Higgs_10} except with $\tan\beta=40$. The green lines (dashed) represent the lower
limit of Eq.\ref{higgsrange} for $m_h$.}        
\label{fig:mh}
\end{center}
\end{figure}

The effects of nonholomorphic parameters particularly $A_t'$ on 
radiative corrections to the CP-even lighter Higgs boson 
mass $m_h$ as enumerated in Eq.\ref{higgs_loop} is particularly prominent for 
smaller $\tan\beta$. 
Keeping this in mind, we choose $\tan\beta=10$ and show 
the extent of variation of $m_h$ due to varying $A_t$ and $A_t'$. 
This is displayed in Fig.\ref{fig:Higgs_10}. 
We assume a 3~\gev~ window in $m_h$ leading to the  
following range\cite{loopcorrection}. 
\begin{equation}
122.1\leqslant~m_h~\leqslant~128.1~\gev.
\label{higgsrange}
\end{equation} 
We note that the above uncertainty that has been widely used arise from renormalization scheme 
related dependencies, scale dependence, problems in computing higher 
order loop corrections up to three loops or the uncertainty 
in the experimental value of top-quark mass 
\footnote{We also remind the reader the additional issue of uncertainty 
of about 2.8 GeV in $m_t^{pole}$ as argued 
in Ref.\cite{Alekhin:2012py}.}.
Our choice of electroweak scale
parameters\footnote{The parameters are given at the scale of 
the geometric mean of the top-squark parameters before mixing. 
The relevant SM parameters used 
are $m_t^{pole}= 173.5~\gev$, 
$m_b^{\overline{MS}} = 4.18~\gev$ and $m_{\tau}=1.77~\gev$.},
 and their ranges, 
that may produce the right amount of radiative corrections to 
Higgs mass are as given below,  
\begin{align}
\label{scan-ranges_10_higgs}
10~\leqslant \mu \leqslant~1000\ \gev, \notag \\
-3000~\leqslant~A_t \leqslant~3000\ \gev, \notag\\
-2000~\leqslant \mu' \leqslant~2000\ \gev, \\
-3000~\leqslant~A_t'~\leqslant~3000\ \gev. \notag
\end{align}
The values of relevant strong sector input like $M_3$, and 
the third generation of scalar mass parameters are fixed at 1.5~\tev~ and 
1~\tev~ respectively that lead to physical states like that of
the gluino or the 
top-squarks to have masses above the LHC limits.
All other trilinear couplings are set to zero.
Finally, without losing any generality we do the analysis for a fixed choice of gaugino masses namely, 
$M_1=150$ \gev, $M_2=250$ \gev.  
We compute the spectrum 
using $\tt SPheno$\cite{spheno_refs} [v.3.3.3 ] while implementing
 the model from $\tt SARAH$\cite{sarah_refs} [v.4.4.4]. The sparticle 
 mass limits are also taken into account\cite{Agashe:2014kda}.  The ranges mentioned in Eq.\ref{scan-ranges_10_higgs} correspond to the results of Sec.\ref{mh_analysis} and \ref{low-energy-b-physics}. 
    
\noindent
In order to probe NHSSM signatures on the Higgs boson mass
we plot both the NHSSM and MSSM specific parameter points in 
Fig.\ref{fig:Higgs_10}.  
The magenta colored points in Fig.\ref{fig:Higgs_10}
correspond to the NHSSM 
scenario where variation due to relevant holomorphic and 
non-holomorphic parameters are 
as referred in Eq.\ref{scan-ranges_10_higgs}. 
We isolate the MSSM specific parameter points in the cyan colored 
region by a choice of $A_t'=\mu'=0$. 
Clearly, focusing on the non-maximal region of $m_h$ and a 
given value of $A_t$ we note that 
the lighter Higgs boson mass may have a $2-3~\gev$~ amount of 
enhancement/decrease, a signature of NHSSM.  
Additionally, compared to MSSM, NHSSM is able to 
provide with correct ranges of 
$m_h$ for a significantly lower value of $|A_t|$.
We must, however, note that for smaller $|A_t|$ less than a TeV or so, the
contribution from $A_t'$ to the radiative corrections to
the Higgs mass is hardly large enough so as to satisfy the lower limit
of Eq.\ref{higgsrange}. 
Needless to mention 
a choice of a heavier third generation of squark would easily 
enhance $m_h$ close to its upper bound of Eq.\ref{higgsrange}.  

\noindent 
Now, we would like to focus on $\tan\beta=40$.  Since the contribution of $A_t'$ 
is suppressed by $\tan\beta$ (Eq.\ref{stopmassmatrix}), 
we expect only a marginal impact on $m_h$.  This is  
evident from the appearance of only a 
small spread around the cyan region of Fig.\ref{fig:Higgs_40}.
Although we have considered $A_b'$ to be vanishing, it may be noted 
that a significantly larger value of $A_b'$ may lead to non-negligible contribution towards $m_h$ via the effect of sbottom loops.
This effect is indeed enhanced  by $\tan\beta$ 
and depends on the off-diagonal quantity 
$X_b$, where $X_b = \{A_b - (\mu + A_b')\tan\beta\}$\footnote {See \cite{Chattopadhyay:2014gfa} and references therein along with references for the NHSSM and the MSSM.}.  Similar contribution arises also from the stau loops 
in the presence of large $A_{\tau}'$ in the large $\tan\beta$ regime.  However for our analysis of Fig.\ref{fig:mh} with $\mu< 1$~\tev, 
we hardly expect any significant contribution to $m_h$ from the  
sbottom loops since the prefactor of $\tan\beta$ in the off-diagonal
sbottom mass matrices is not too large with the given range of
parameter regions considered in this analysis. This is also true for
stau loops. 
Apart from the effect of NH trilinear parameters particularly $A_t'$ 
on $m_h$ via top-squarks in the loops,
we must remember that the other NH parameter $\mu'$ may play an
important role via the chargino loop contributions to the
Higgs boson mass\cite{Haber:1990aw,Ibrahim:2000qj}. 
The latter contributions that are intrinsically negative
are essentially independent of $\tan\beta$\cite{Haber:1990aw,Ibrahim:2000qj}. As a result, in the region near $A_t=0$
in Fig.\ref{fig:Higgs_40} where $A_t'$ is not able to influence on $X_t$
because of suppression via $\tan\beta$, 
we find a spread of the magenta points toward the
smaller direction of $m_h$.
That the effect does not depend on $\tan\beta$ is manifested
in the similar region of Fig.\ref{fig:Higgs_10}. Thus, larger values of
$\mu-\mu'$ may cause a decrease in $m_h$ for both values of
$\tan\beta$ as used in the figure.  
Although enhancement of $m_h$ due to  
NH parameters is not very significant in the 
large $\tan\beta$ limit, the impact of the above parameters on 
particularly the low energy phenomenological constraints like 
${\rm Br}(B \rightarrow  X_s + \gamma)$ is extremely important
as we will see in Sec.\ref{low-energy-b-physics}.
\subsection{Effects of non-holomorphic parameters on SUSY contributions to $\bsg$ and $\bsmumu$}
\label{low-energy-b-physics} 
In this subsection, we would like to discuss the results of including 
the constraints of ${\rm Br}(B \rightarrow X_s +  \gamma)$ 
and ${\rm Br}(B_s \rightarrow \mu^+ \mu^-)$ on the 
NHSSM parameter space. 
The experimental limits 
on ${\rm Br}(B \rightarrow X_s + \gamma)$ at $3\sigma$ level reads\cite{Amhis:2012bh}
\begin{align}
2.77\times 10^{-4}~\leqslant~{\rm Br}(B \rightarrow X_s + \gamma)~\leqslant~4.09\times 10^{-4}.
\label{bphysics_constraints1}
\end{align}
On the other hand, the 
recent constraints from ${\rm Br}(B_s \rightarrow \mu^+ \mu^-)$ 
at $3\sigma$ level results
into\cite{Chatrchyan:2013bka,Aaij:2013aka,CMSandLHCbCollaborations:2013pla} 
\begin{align}
0.8 \times 10^{-9}~\leqslant~{\rm Br}(B_s \rightarrow \mu^+ \mu^-)~\leqslant~5\times 10^{-9}. 
\label{bphysics_constraints2}
\end{align}

\begin{figure}[htb]
     \begin{center}
        \subfigure[]{%
            \label{fig:mh_bphy_10}
            \includegraphics[width=0.51\textwidth]{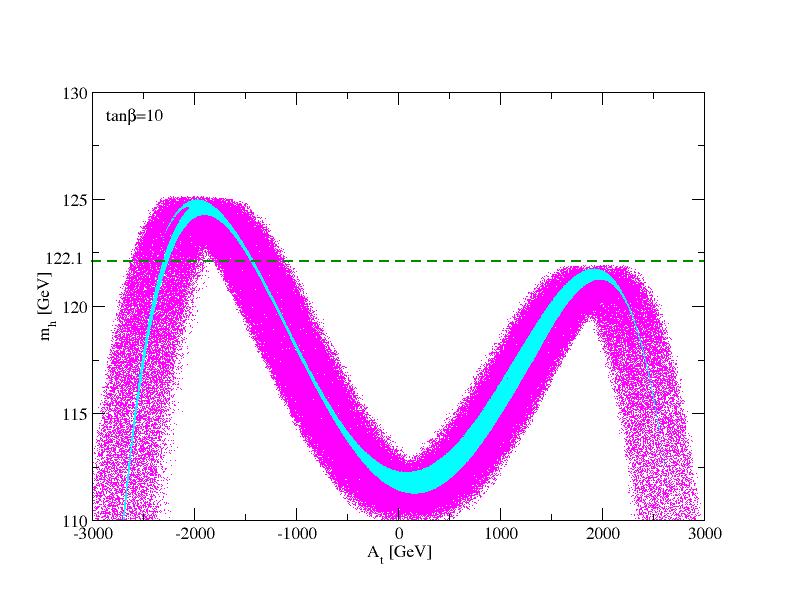}
        }%
        \subfigure[]{%
           \label{fig:mh_bphys_40}
           \includegraphics[width=0.51\textwidth]{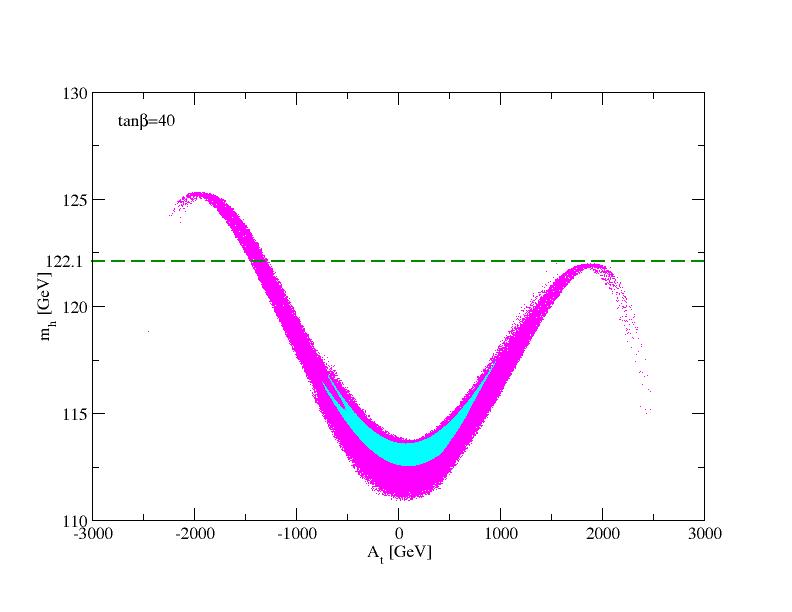}
        }
    \caption{\it The variation of $m_h$ against $A_t$ for the 
scanning ranges of Eq.\ref{scan-ranges_10_higgs}, with $\tan\beta=10$ and $\tan\beta=40$. 
The magenta and cyan colored
regions correspond to 
NHSSM and MSSM respectively. The green lines (dashed) represent the lower
limit of Eq.\ref{higgsrange} for $m_h$. Furthermore, we impose the 
constraints from ${\rm Br}(B \rightarrow X_s + \gamma)$ and 
${\rm Br}(B_s \rightarrow \mu^+ \mu^-)$ on the 
resulting spectrum. Clearly, for $\tan\beta=40$
a large region of parameter space in MSSM with large $A_t$ 
is excluded by these constraints.
However, NHSSM essentially recovers the     
large $A_t$ regions consistent with the Higgs mass as well as  the B-physics 
constraints.}
\label{fig:mh_bphy}
\end{center}
\end{figure}
\noindent
Fig:~\ref{fig:mh_bphy_10} for $\tan\beta=10$ shows the effect of imposing 
the B-physics constraints on the parameter space of 
MSSM and NHSSM that are displayed  
with different colors. 
Both the MSSM (cyan) and NHSSM (magenta) parameter regions of Fig:~\ref{fig:mh_bphy_10} hardly show any change when B-physics constraints are imposed 
in comparison to Fig.\ref{fig:Higgs_10} where the same were not included. 
Since $\tan\beta$ is not large one does not 
expect any significant degree of change in the MSSM parameter space 
when $\bsg$ constraint is applied because the SUSY contribution of the same 
approximately scales with 
$\tan\beta$\cite{Haisch:2012re}\footnote{For further discussion see \cite{Chattopadhyay:2014gfa,Bhattacherjee:2013vga} and references therein.}. The scaling 
behavior also holds good in NHSSM.  

Fig.\ref{fig:mh_bphys_40} for $\tan\beta=40$, shows that 
the constraints of 
Eqs.\ref{bphysics_constraints1} \& \ref{bphysics_constraints2}
exclude a large amount of MSSM parameter region (cyan) 
when $|A_t|$ is large. 
This is indeed expected  
with the scaling behavior with respect to $\tan\beta$ as mentioned 
above in regard to the $\bsg$ constraint.
The region with large $A_t$ along with 
$\mu A_t<0$ ($\mu$ is scanned over positive value as in Eq.\ref{scan-ranges_10_higgs}) is discarded via 
the lower bound of Eq~\ref{bphysics_constraints1} whereas the region with $\mu A_t>0$ is disallowed 
via the upper bound of Eq~\ref{bphysics_constraints1}\footnote{Specifically, see 
Eq.2.28 and 2.30 of Ref.\cite{Haisch:2012re} in relation to 
the ${\rm Br}(B \rightarrow X_s +\gamma)$ constraint. 
Refs.\cite{bsgammaextra,bsmumuextra} may be seen for similar other 
analyses in MSSM.}.  Thus, a large $|A_t|$ regions become unavailable in MSSM which in turn causes $m_h$ to go below the lower limit of Eq.\ref{higgsrange}.  Certainly, $m_h$ can be increased via increasing the third generation of scalar mass that would enhance the Higgs mass radiative corrections.    
 
In contrast to MSSM, we find that the magenta region corresponding 
to NHSSM includes parameter points that satisfy the Higgs mass bounds in addition to the B-physics constraints.  
Thus large values of $A_t$ (with preference to negative region) 
correspond to valid parameter zones simply 
because of the fact that the role played by $A_t$
 is effectively replaced by $A_t-A_t'\cot\beta$ in NHSSM 
(see Sec.\ref{bphysics}). 
Thus a scan over $A_t'$ even for a large value of $\tan\beta$ in NHSSM is able to accommodate appreciably 
large values of $|A_t|$ consistent with the Higgs mass as well as B-physics 
constraints.  
\subsection{Electroweak fine-tuning and higgsino dark matter}
\label{finetuninganalysis}
Typically a higgsino dominated dark matter with a mass around a few hundred 
GeV produces extremely large annihilation cross section.  Apart from the LSP pair annihilation there is a substantial amount of $\lspone - \tilde{\chi_1}^{\pm}$ coannihilation.   
A larger higgsino content in a primarily bino dominated $\lspone$ 
such as what one obtains in the focus point\cite{FP1,FP2}/hyperbolic branch scenario\cite{ewft_earlier_works} 
may produce the right relic abundance satisfying the experimental 
constraint of DM relic density. However, this is highly constrained by 
the direct detection of DM experiments like LUX \cite{Akerib:2013tjd}.  
In MSSM, a highly higgsino dominated LSP satisfies the DM relic density 
limits as given by Eq.\ref{omega_eqn} from PLANCK \cite{Ade:2013zuv} data 
for an LSP mass of $\sim 1$~TeV\cite{ArkaniHamed:2006mb,Chakraborti:2014fha}. 
\begin{align}
0.092\leqslant\varOmega_{\lspone}h^2\leqslant~0.138.
\label{omega_eqn}
\end{align}
Certainly, with a 1~TeV higgsino mass the electroweak fine-tuning estimate 
$\Delta_{Total}$ of Eq.\ref{formulaewft} becomes generally large\footnote{See however Ref.\cite{Kowalska:2014hza} where the authors considered specific GUT scenarios with non-universal gaugino masses.}. 
One, on the other hand, is able to isolate significantly the 
EWFT measure from LSP mass in NHSSM   
simply because of the fact that the Higgs potential (Eq.\ref{higgs_potential})
does not depend on the nonholomorphic Higgsino parameter $\mu'$, whereas 
the higgsino content of the LSP is determined via  
the difference of $\mu$ and $\mu'$.

\begin{figure}[htb]
     \begin{center}
        \subfigure[]{%
            \label{fig:ewft_10}
            \includegraphics[width=0.51\textwidth]{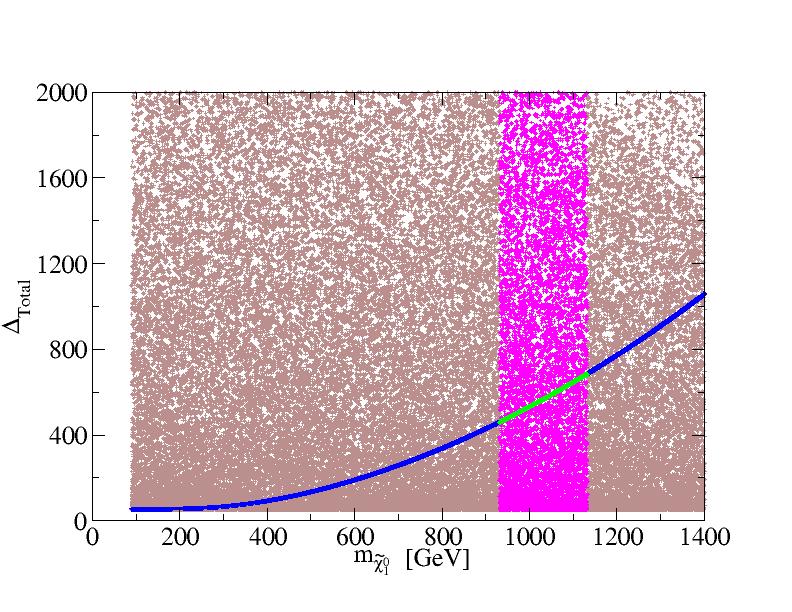}
        }%
        \subfigure[]{%
           \label{fig:ewft_40}
           \includegraphics[width=0.51\textwidth]{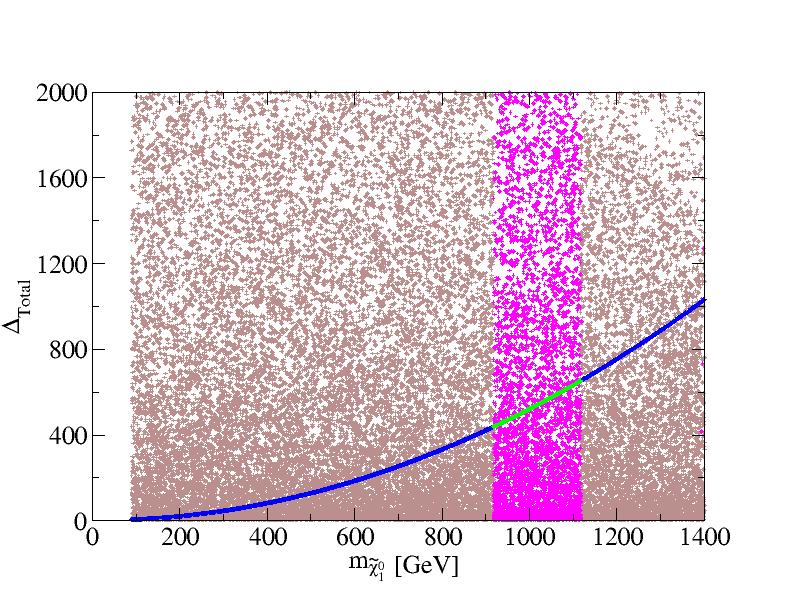}
        }
    \caption{\it The variation of $\Delta_{Total}$ against $\mlspone$ 
for the scanning ranges of Eqs.\ref{scan-ranges_higgsino_dark_matter} and \ref{At_Atp_scan}, 
with $\tan\beta=10$ and $\tan\beta=40$.  
The NHSSM and MSSM are shown in brown and blue colors respectively. The
scan does not include the Higgs mass range of Eq.\ref{higgsrange}. The relic density limits of Eq.\ref{omega_eqn} are satisfied in the vertical strip shown in magenta or green corresponding to NHSSM and MSSM cases. It is evident from the figures that EWFT can be significantly lower in NHSSM in a region with higgsino-like LSP providing the required relic
abundance.}
\label{fig:ewft}
\end{center}
\end{figure}

\noindent
We now scan the NHSSM parameter space focusing on a higgsino 
dominated LSP. We select a fixed value of 3 TeV for all 
the following masses namely $M_1$,$M_2$ and $m_A$ along with 
all the squark and slepton mass parameters in a scenario of 
varying $\mu-\mu'$ as mentioned below. 
\begin{align}
\label{scan-ranges_higgsino_dark_matter}
-3~\tev ~\leqslant \mu \leqslant~3~\tev, \notag\\
-3~\tev ~\leqslant \mu' \leqslant~3~\tev. 
\end{align}
We note that the LEP limit on lighter chargino mass is translated in NHSSM as $|\mu-\mu'| \gsim 100$~GeV and we probe an LSP mass zone up to 
$1.5~\tev$~ so that $\lspone$ remains sufficiently higgsino dominated in its 
composition whereas we do not include the s-channel $A$-boson annihilation region.  Furthermore, in order to have the Higgs 
mass in the correct range, we vary the 
trilinear parameters $A_t$ and $A_t'$ as given below\footnote{However, while doing a generic study on fine-tuning we do not impose any explicit
  constraints like Higgs mass (Eq.\ref{higgsrange}) or
  B-physics limits
  (Eqs.\ref{bphysics_constraints1},\ref{bphysics_constraints2}).}.    
\begin{align}
\label{At_Atp_scan}
-3~\tev ~\leqslant~A_t~\leqslant~3~\tev , \notag\\
-3~\tev ~\leqslant~A_t'~\leqslant~3~\tev .
\end{align}
The parameter ranges of Eq.\ref{scan-ranges_higgsino_dark_matter} and \ref{At_Atp_scan} apply to the discussion of the present subsection only.  
In Fig.\ref{fig:ewft} we plot $\Delta_{Total}$ (Eq.\ref{formulaewft}) vs $\mlspone$
for MSSM and NHSSM cases drawn in blue and brown colors respectively. 
The MSSM part of the analysis corresponds to vanishing $\mu'$ and $A_t'$ 
and appears as a thin blue line in the middle.  The relic density limits are satisfied in the vertical strip shown in magenta or green corresponding to NHSSM and MSSM cases.  The lighter chargino mass limit
disallows the LSP mass to go below $100~\gev$~(white region in the left).  Fig.\ref{fig:ewft_10} shows that $\Delta_{Total}$ for NHSSM 
can either be larger or smaller than the MSSM specific values. The larger $\Delta_{Total}$  
region, of course, occurs when $|\mu|$ is larger while the lower region 
corresponds to smaller $|\mu|$. The LSP mass is essentially  same  
as $|\mu-\mu'|$.  Fig.\ref{fig:ewft_40} shows a similar result for $\tan\beta=40$.
The lowest value of 
$\Delta_{Total}$ satisfying the DM relic density constraint of 
Eq.\ref{omega_eqn} for MSSM is about 500, whereas the same for NHSSM for
$\tan\beta=10$ is about 50 rather than being vanishingly small\footnote{We essentially agree with the analysis of Ref.\cite{Ross:2016pml}.}. On the
other hand, $\Delta_{Total}$ for $\tan\beta=40$ 
can indeed approach zero for vanishingly small $\mu$.
The above difference of small $\mu$ behavior of $\Delta_{Total}$ is indeed
consistent with the discussion of Sec.~\ref{ewft_theory}.
We have used {\tt micrOMEGAs}\cite{micromegas_ref} for relic 
density computation.  

We note that a higgsino type of LSP generally satisfies the LUX data\cite{Chakraborti:2014fha}
and it may be probed via XENON1T \cite{Aprile:2012zx}.
\subsection{Constraint from muon $g-2$ in relation to large $\amususy$ in NHSSM}
\label{gmin2analysis}
In this subsection, we would like to demonstrate a novel signature of NHSSM on $\amususy$ by showing the degree of influence of the 
NH trilinear parameter $A_{\mu}'$.  We would particularly stress on the fact 
that even a small value of $A_{\mu}'$ like 50~GeV can cause a tremendous 
change in $\amususy$ when compared to the corresponding MSSM scenario. 
Clearly, this is possible when the neutralino-smuon loops dominate 
over the chargino-sneutrino loops in their contribution to $\amususy$. Keeping 
this in mind we study the effect of Muon $g-2$ constraint on the $m_{{\tilde \mu}_1} - \mlspone$ plane while selecting a low range for $M_1$ 
satisfying $ M_1 < \mu < M_2$, 
corresponding to fixed values of $\mu$ and $M_2$ namely,  $\mu=500$~GeV and $M_2=1500$~GeV.  The scanning of NHSSM parameter space with $\mu'=0$ is considered in a background of fixed squark and stau masses set 
at 1~TeV, while choosing  
$A_t=-1.5$~TeV with all other trilinear parameters being set to zero.  The range of variation considered for $M_1$ and 
the first two generation of slepton masses $M_{\tilde l}$ are as follows. 
\begin{eqnarray}
100 ~{\rm GeV} < M_1 < 400 ~{\rm GeV} ,\nonumber \\
100 ~{\rm GeV}  < M_{\tilde l} < 1000 ~{\rm GeV}.
\label{muon_scan}
\end{eqnarray}

\begin{figure}[htb]
     \begin{center}
        \subfigure[]{%
            \label{fig:muon_10_0}
            \includegraphics[width=0.51\textwidth]{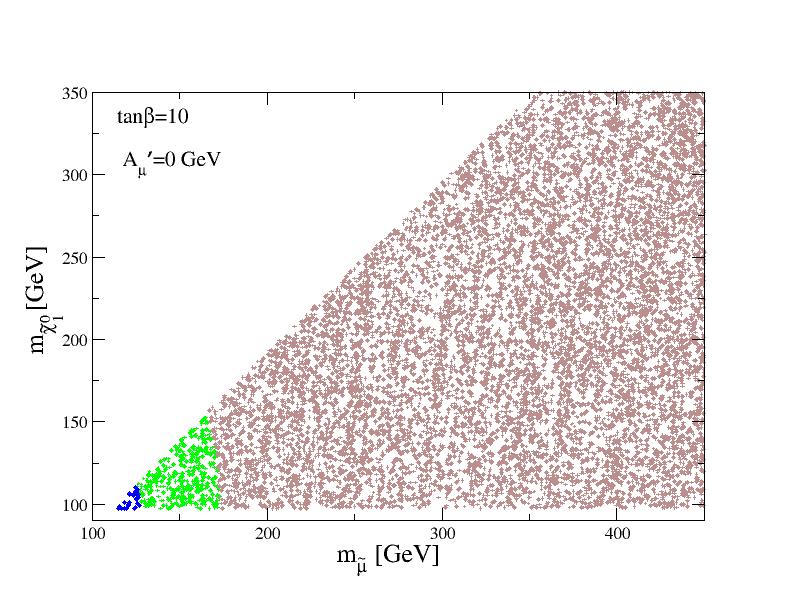}
        }%
        \subfigure[]{%
           \label{fig:muon_40_0}
           \includegraphics[width=0.51\textwidth]{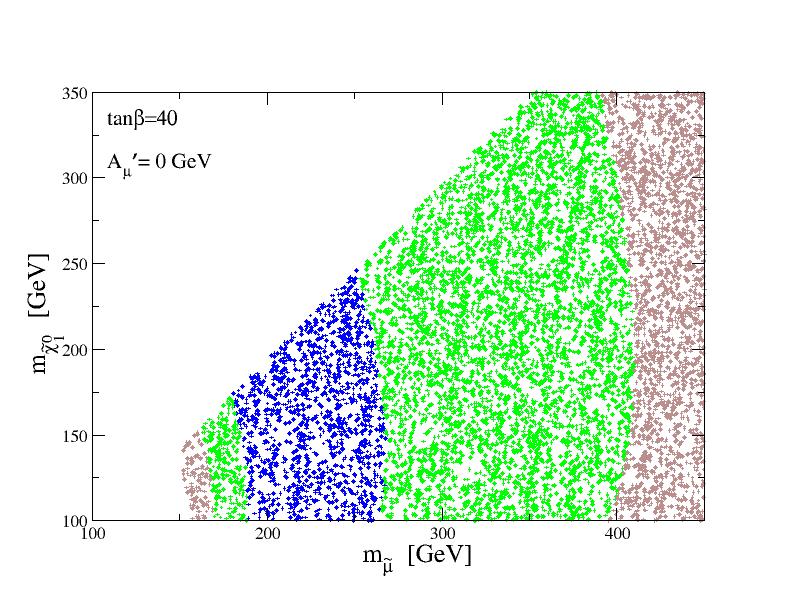}
        }
    \caption{\it Scattered plot of $\mlspone$ against $m_{{\tilde \mu}_1}$ 
      for $\tan\beta=10$ and $40$ in MSSM for $\mu=500$~GeV and 
$M_2=1500$~GeV. All the squark and stau mass
      parameters are set at 1~TeV along with choosing vanishing MSSM
      trilinear couplings except $A_t$ which  is set at $-1.5$~TeV,
      favorable to have a correct Higgs boson mass. The
      Higgs mass constraint of Eq.\ref{higgsrange} is however not imposed.
      The blue, green and brown regions correspond to satisfying the 
muon $g-2$ constraint within $1\sigma$, $2\sigma$ and $3\sigma$
limits respectively.
}
\label{fig:muon_0}
\end{center}
\end{figure}

Fig.\ref{fig:muon_0} shows the parameter points in the $m_{{\tilde \mu}_1} - \mlspone$ plane corresponding to the MSSM scenario for $\tan\beta=10$ and $40$ where 
we isolate the degree of satisfying the Muon $g-2$ constraint at 
1$\sigma$, 2$\sigma$ and 3$\sigma$ levels as shown in blue, green and 
brown  colors respectively. The upper limits of 
$m_{{\tilde \mu}_1}$ at 1$\sigma$ level in Fig.\ref{fig:muon_10_0} and 
Fig.\ref{fig:muon_40_0} are about 125~GeV and 260~GeV respectively. 
Thus one requires very light ${\tilde{\mu}}_1$ in order
to have Muon $g-2$ within $1\sigma$ limits. This will drastically change 
in NHSSM as we will see in the following figures.      

\begin{figure}[htb]
     \begin{center}
        \subfigure[]{%
            \label{fig:muon_10_300}
            \includegraphics[width=0.51\textwidth]{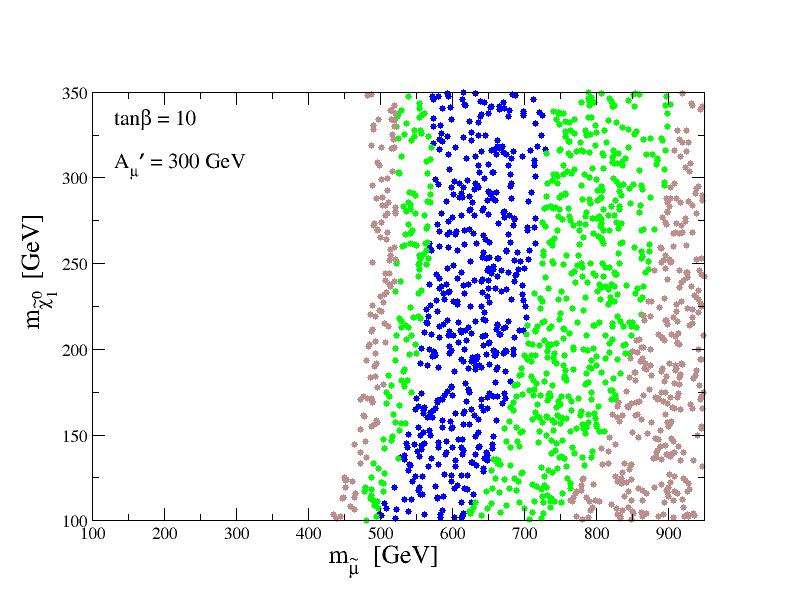}
        }%
        \subfigure[]{%
           \label{fig:muon_40_50}
           \includegraphics[width=0.51\textwidth]{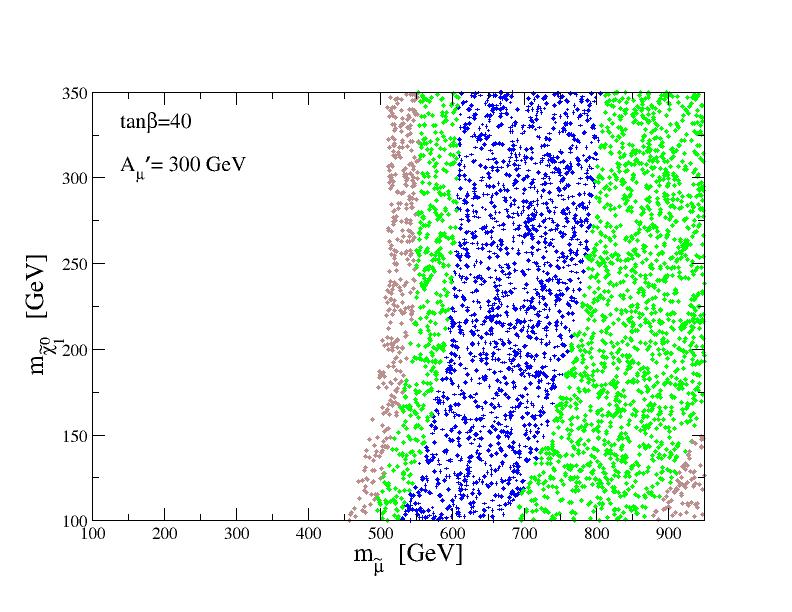}
        }
    \caption{\it Scattered plot of $\mlspone$ against $m_{{\tilde \mu}_1}$ 
in NHSSM for the scanning ranges of 
Eq.\ref{muon_scan}, with $\tan\beta=10$ and $40$ for 
$A_{\mu}'=300$~GeV. The color scheme along with the relevant MSSM
parameters are same as in Fig.\ref{fig:muon_0}.}
\label{fig:muon_300}
\end{center}
\end{figure}

\noindent
Fig.\ref{fig:muon_50} for NHSSM shows the parameter 
points in the same plane corresponding 
to $\tan\beta=10$ and $40$ for a fixed value of $A_{\mu}'=50$~GeV. 
Even with such a small value of $A_{\mu}'$ we see that the upper limits of 
$m_{{\tilde \mu}_1}$ at 1$\sigma$ level jumping to 420~GeV and 500~GeV 
respectively. Clearly, the contribution to $\amususy$ is visibly  
substantial when we compare the above with the results of the 
MSSM case of Fig.\ref{fig:muon_0}.  The analysis is further extended for $A_{\mu}'=300$~GeV in 
Fig.\ref{fig:muon_300}. The same upper limits of $m_{{\tilde \mu}_1}$ are 
now 750~GeV and 800~GeV, almost impossible to reach within MSSM whatsoever 
while assuming the dominating loops to involve neutralinos rather than 
charginos.

\noindent
Thus NHSSM can easily accommodate the stringent muon $g-2$ constraint even with a 
small amount of NH trilinear coupling $A_{\mu}'$ by allowing 
larger smuon masses. Apart from the above in relation to the effect of 
the combined constraints from  B-physics and Muon $g-2$, one finds that 
NHSSM can accommodate the large $\tan\beta$ regimes (that naturally 
increases $\amususy$) easily in comparison 
with the MSSM scenario.  Table~\ref{table-benchmark3} compares MSSM and NHSSM spectra for two benchmark points in detail.  Both the points for NHSSM satisfy all the relevant constraints, whereas the corresponding MSSM points do not necessarily satisfy the same.
\begin{figure}[htb]
     \begin{center}
        \subfigure[]{%
            \label{fig:muon_10_50}
            \includegraphics[width=0.51\textwidth]{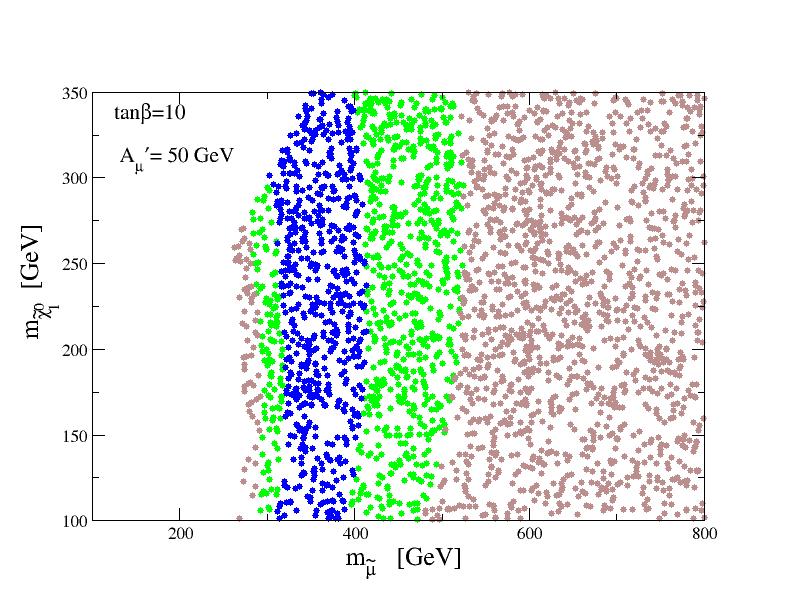}
        }%
        \subfigure[]{%
           \label{fig:muon_40_50}
           \includegraphics[width=0.51\textwidth]{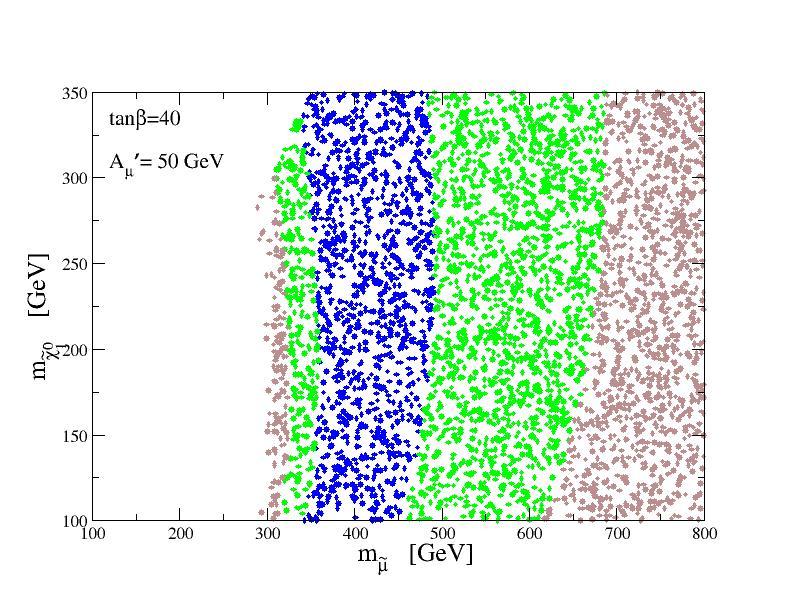}
        }
    \caption{\it Scattered plot of $\mlspone$ against $m_{{\tilde \mu}_1}$ 
in NHSSM for the scanning ranges of 
Eq.\ref{muon_scan}, with $\tan\beta=10$ and $40$ for 
$A_{\mu}'=50$~GeV. The color scheme along with the relevant MSSM
parameters are same as in Fig.\ref{fig:muon_0}. 
}
\label{fig:muon_50}
\end{center}
\end{figure}

\begin{table}
  \caption{\it Benchmark points for NHSSM. Masses are shown in GeV. Only the two NHSSM benchmark points shown satisfy the phenomenological
    constraint of Higgs mass, dark matter relic density along with
    direct detection cross section, muon anomaly, $\bsg$ and $\bsmumu$. The associated MSSM points are only given for comparison and do not necessarily satisfy all the above constraints.}
	\centering
	\begin{tabular}{|c|c|c|c|c|}
		\hline\hline 
		Parameters  &  MSSM &  NHSSM  & MSSM &  NHSSM \\ [0.5ex]
		\hline
		$m_{1,2,3}$  & {472, 1500, 1450} &   {472, 1500, 1450} &  {243, 250, 1450} & {243, 250, 1450}\\
		$m_{\tilde{Q}_{3}}/ m_{\tilde{U}_{3}}/ m_{\tilde{D}_{3}}$  & { 1000} & { 1000} & { 1000} & {1000}\\
		$m_{\tilde{Q}_{2}}/ m_{\tilde{U}_{2}}/ m_{\tilde{D}_{2}}$  & 1000 & 1000 & 1000 & {1000}\\
		$m_{\tilde{Q}_{1}}/ m_{\tilde{U}_{1}}/ m_{\tilde{D}_{1}}$ & 1000 & 1000 & 1000 & {1000}\\
		$m_{\tilde{L}_{3}}/m_{\tilde{E}_{3}}$ & 2236 & 2236 & 1000 & 1000\\
		$m_{\tilde{L}_{2}}/m_{\tilde{E}_{2}}$ & 592 & 592 & 500& 500 \\
		$m_{\tilde{L}_{1}}/ m_{\tilde{E}_{1}}$ & 592 & 592 & 500 & 500\\
		$A_{t},A_{b},A_{\tau}$  & -1500, 0, 0 & -1500, 0, 0 & -1368.1, 0, 0 & -1368.1, 0, 0  \\
		$A_{t}',A_{\mu}',A_{\tau}'$  & 0, 0, 0 & 2234, 169, 0 & 0, 0, 0 & 3000, 200, 0 \\
		$\tan\beta$ & {\bf 10} & {\bf 10} & {\bf 40} & {\bf 40} \\
		$\mu$  & 500 &  500 & 390.8 & 390.8\\
		$\mu'$  & 0 &  -175  & 0 & 1655.5\\
		$m_A$  & 1000 & 1000 & 1000 & 1000 \\
		\hline 
		$m_{\tilde g}$ & 1438.9 & 1439.1 & 1438.9 & 1438.9\\
		$m_{\tilde t_1},m_{\tilde t_2}$& 894.4, 1151.2 & 865.5, 1154.9 & 907.8, 1137.5 & 903.4, 1141.4 \\
		$m_{\tilde b_1},m_{\tilde b_2}$  & 1032.4, 1046.2 & 1026.3, 1045.1 & 1013.8, 1051.2 & 1017.7, 1056.5\\
		$m_{\tilde \mu_L}, m_{\tilde {\nu_{\mu}}}$ & 596.4, 596.3 &  573.5, 595.9 & 502.0, 497.1 & 465.8, 496.3 \\
		$m_{{\tilde \tau}_1},m_{\tilde {\nu_\tau}}$  & 2237.1, 2238.5 &  2237.1, 2238.5 & 985.4, 997.2 & 988.5, 998.8\\
		$m_{{\tilde \chi_1}^{\pm}},m_{{\tilde \chi_2}^{\pm}}$  & 504.2, 1483.6 & 677.6, 1484.7 & 244.6, 421.0 & 262.3, 1255.2 \\
		$m_{{\tilde \chi_1}^0},m_{{\tilde \chi_2}^0}$ & 448.6, 509.0 &  464.0, 680.6 & 231.3, 249.9 & 240.9, 262.1 \\
		$m_{{\tilde \chi_3}^0},m_{{\tilde \chi_4}^0}$ & 522.6, 1483.5 & 683.2, 1484.7 & 400.7, 421.0 & 1253.3, 1253.7\\
		$m_{H^{\pm}}$ & 1011.9 & 1005.8 & 955.7 & 1011.6\\
		$m_H,m_h$ & 1008.1, 121.4 & 984.8, 122.8 & 948.0, 122.4 & 990.2, 122.8\\
		\hline
		${\rm Br}(B \rightarrow X_s + \gamma)$ &$3.00\times10^{-4}$ & $3.01\times10^{-4}$ & $2.01\times10^{-4}$ & $4.05\times10^{-4}$ \\
		${\rm Br}(B_s \rightarrow \mu^+ \mu^-)$  & $3.40\times10^{-9}$ & $3.45\times10^{-9}$ & $5.06\times10^{-9}$ & $1.65\times10^{-9}$\\
		$a_{\mu}$ & $1.94\times10^{-10}$ & $22.3\times10^{-10}$ & $34.8\times10^{-10}$ & $35.8\times10^{-10}$ \\
		\hline
		$\varOmega_{\lspone}h^2$ & 0.035 & 0.095 & 0.0114 & 0.122\\
		$\rm  {\sigmachionesip}$ in pb & $4.01\times10^{-9}$ & $3.47\times10^{-10}$ & $6.79\times10^{-9}$ & $3.15\times10^{-12}$\\
		\hline
	\end{tabular}
	\label{table-benchmark3}
\end{table}
Finally, we will comment on the possible effect of considering a negative 
sign of $\mu$. In regard to Sec.\ref{mh_analysis} 
the part involving the radiative corrections to 
Higgs boson mass is controlled by 
$X_t=A_t-(\mu+A_t')\cot\beta$. Hence for large $\tan\beta$ the 
contribution coming from the part involving $\mu$ and $A_t'$ is 
suppressed.  On the other hand, for smaller values of $\tan\beta ~(=10)$ 
 the above contribution is relatively 
larger. However, since $A_t$ as well as $A_t'$ both are scanned with a 
larger range ($-$3 TeV to 3 TeV) than what is used for 
$\mu$ (up to 1 TeV), a negative $\mu$ would not lead to a much different 
result in the given plane of the figures.  
The conclusion is also similar for 
Sec.\ref{low-energy-b-physics} where both signs of 
$A_t$ and $A_t'$ are used with a larger range than what is used for $\mu$. 
For fine-tuning estimate given in Sec.\ref{finetuninganalysis}, the result 
would be essentially unchanged since the measure 
depends on $\mu$ quadratically unless $\tan\beta$ is small and
$\mu$ is vanishingly small.  
In regard to $\amususy$ where a fixed sign of gaugino masses are 
used in our analysis of Sec.\ref{gmin2analysis}, use of a negative $\mu$ would 
require appropriate values of $A_\mu'$ so as to satisfy 
the experimental data. This is via the contribution from the 
neutralino-smuon loops to $\amususy$. 

\clearpage
\section{Conclusion}
\label{conclusion}
In MSSM the superpotential is a holomorphic function of superfields and 
one considers soft SUSY breaking terms that are also holomorphic function of 
fields. However a SUSY theory 
devoid of an SM gauge singlet allows 
non-holomorphic soft SUSY breaking terms in the Lagrangian, 
and this has been used in various beyond the MSSM scenarios  
analyzing neutrino physics, leptogenesis, CP violation etc.
In this analysis, we focus on the relevant phenomenological constraints 
while considering non-holomorphic soft SUSY breaking terms in a beyond the 
MSSM scenario with identical particle content as that of MSSM.
Our work on Non-Holomorphic Supersymmetric Standard Model (NHSSM)   
uses electroweak scale input parameters similar to what is considered in 
the pMSSM model. This is unlike the previous analyses where only the 
non-holomorphic parameters 
were given at the electroweak scale while other soft parameters belonged 
to the grand unification scale or all the input parameters 
were given at the aforesaid scale.

\noindent 
We particularly analyze NHSSM specific effects on the Higgs mass radiative 
corrections, electroweak fine-tuning, electroweakino spectra, the constraint due to $\bsg$ and $\bsmumu$
and the novel signature of NHSSM that enhances $\amususy$ so that 
it can easily accommodate the Muon $g-2$ limits even for larger 
smuon masses or small $\tan\beta$.    
In the context of radiative corrections to the Higgs boson mass 
it is seen that for $\tan\beta=10$ and a given value of $A_t$, the 
variation of NHSSM parameters, particularly due to $A_t'$, the trilinear 
nonholomorphic parameter may easily cause a change in $m_h$ by 
2-3 TeV in either direction positive or negative. 
In other words, NHSSM does not necessarily require large values of 
$|A_t|$ in order to produce the right amount of radiative corrections 
to Higgs mass,  since both $A_t$ and $A_t'$ contribute toward the corrections. 
The radiative contributions to $m_h$ due to $A_t'$ is 
suppressed by $\tan\beta$ leading to a quite small effect for 
$\tan\beta=40$. However, for large $\tan\beta$ NHSSM makes 
parameter space with large $|A_t|$ to become valid via its effects on $\bsg$. 
We note that with squark and gluino masses assuming just above the LHC limits and $\tan\beta=40$ 
the constraints from $\bsg$ in MSSM eliminates the 
large $|A_t|$ zones altogether so that $m_h$ goes below 122.1~GeV, 
the lower limit of Higgs mass considered in the analysis. 
The non-holomorphic terms in NHSSM through their 
contributions toward $\bsg$ allows a significant part of the 
large $|A_t|$ region to become valid.  

\noindent
It is known that a higgsino type of LSP in MSSM 
that is supposed to satisfy the 
PLANCK data on DM relic density has a mass of around 1~TeV. In pMSSM this 
obviously increases the electroweak fine-tuning due to the sufficiently 
large value of $\mu$.  In contrast, NHSSM is able to produce a  
drastic reduction of the electroweak fine-tuning 
measure even for such a large mass of higgsino.  The dependence of electroweak fine-tuning on 
$\mu$ rather than $\mu'$, the bilinear Higgs nonholomorphic parameter whereas 
the fact that electroweakino masses are related to the difference of 
$\mu$ and $\mu'$ indeed isolates the two sectors\footnote{As mentioned before, we agree with the result of the recent analysis of ref.\cite{Ross:2016pml} in this regard.}.  The electroweak 
fine-tuning can either decrease 
or increase depending on the relative contributions of $\mu$ and $\mu'$ to 
the difference $\mu-\mu'$. 

\noindent  
Regarding the $\gmin2$ constraint, NHSSM is able to 
significantly enhance $\amususy$ even 
for a small 
$\tan\beta$ via a small value of the associated 
trilinear coupling parameter $A_\mu'$. This is true even for a 
significantly large smuon mass $m_{\tilde {\mu_1}}$.  This is indeed a very novel feature of NHSSM.  Just a small amount $A_\mu'$ like 50~GeV even for a small $\tan\beta$ $(=10)$ 
may significantly alter the MSSM predictions on lighter smuon 
mass that would satisfy the $\gmin2$ constraint 
at 1$\sigma$ level . This has the 
potential to cause a significant change in predictions  
involving SUSY models in general. 
\section{Acknowledgment}
A.D. would like to thank the Council of Scientific and Industrial Research, Government of
India and Indian Association for the Cultivation of Science 
for support and computational facilities. UC would like to thankfully 
acknowledge a correspondence with Florian Staub.

\end{document}